\documentclass[10pt,british,sigplan,10pt,screen]{acmart}
\pdfoutput=1

\usepackage[LGR,T1]{fontenc}
\usepackage[latin9]{inputenc}
\usepackage{color}
\usepackage{cprotect}
\usepackage{refstyle}
\usepackage{mathtools}
\usepackage{algorithm2e}
\usepackage{amssymb}
\usepackage{xargs}[2008/03/08]

\makeatletter

\global\long\def\emptytex{}%
\global\long\def\b#1{\textcolor{black}{#1}}%
\global\long\def\k#1{\textcolor{blue}{\texttt{#1}}}%
\global\long\def\t#1{\textcolor{teal}{#1}}%
\global\long\def\f#1{\textcolor{purple}{#1}}%
\global\long\def\l#1{\textcolor{brown}{#1}}%
\global\long\def\v#1{\textcolor{orange}{#1}}%
\global\long\def\M#1{\textcolor{green}{#1}}%
\global\long\def\c#1{\b{\texttt{#1}}}%
\global\long\def\ck#1{\c{\k{#1}}}%
\global\long\def\ct#1{\c{\t{#1}}}%
\global\long\def\cf#1{\c{\f{#1}}}%
\global\long\def\cv#1{\c{\v{#1}}}%
\global\long\def\cM#1{\c{\M{#1}}}%
\global\long\def\cl#1{\c{\l{#1}}}%
\global\long\def\cb#1{\c{\b{#1}}}%
\global\long\def\mb#1{\ensuremath{\b{#1}}}%
\global\long\def\mt#1{\ensuremath{\t{#1}}}%
\global\long\def\mf#1{\ensuremath{\f{#1}}}%
\global\long\def\mv#1{\ensuremath{\v{#1}}}%
\global\long\def\*{\cf *}%
\global\long\def\ob#1{\colorlet{orig}{.}\b{\ensuremath{\overline{\textcolor{orig}{#1}}}}}%
\global\long\def\sop#1{{}#1{}}%
\global\long\def\up#1{\left[#1\right]}%
\global\long\def\fixec{\raisebox{0.22ex}{:}}%

\global\long\def\cco{\sop{\mathrel{\cb{\fixec\fixec}}}}%
\global\long\def\eq{\cb =}%
\global\long\def\p#1{\texttt{(}\mb{\b{#1}}\texttt{)}}%
\global\long\def\br#1{\texttt{\{}\mb{#1}\texttt{\}}}%
\global\long\def\sq#1{\c [#1\c ]}%
\global\long\def\sceq{\cb{\fixec=}}%
\global\long\def\ceq{\sop{\mathrel{\sceq}}}%

\global\long\def\sc{\sop{\mathpunct{\cb ;}}}%
\global\long\def\ddd{\hbox to\fontcharwd\font`x{\hss.\hss\hss.\hss\hss.\hss}}%
\global\long\def\GrowRule#1{\GrowBox{\RuleVSpace}{\ensuremath{#1}}}%

\global\long\def\hsep{\quad}%

\global\long\def\name{\text{M\textmu l}}%
\global\long\def\newk{\text{\ensuremath{\ck{new}}}}%
\global\long\def\defk{\ck{def}}%
\newcommandx\gt[1][usedefault, addprefix=\global, 1=\emptytex]{\mv{t#1}}%
\newcommandx\x[1][usedefault, addprefix=\global, 1=\emptytex]{\mv{x#1}}%
\newcommandx\y[1][usedefault, addprefix=\global, 1=\emptytex]{\mv{y#1}}%
\renewcommandx\y[1][usedefault, addprefix=\global, 1=\emptytex]{\mv{y#1}}%
\newcommandx\CC[1][usedefault, addprefix=\global, 1=\emptytex]{\mt{C#1}}%
\newcommandx\m[1][usedefault, addprefix=\global, 1=\emptytex]{\mf{\mu#1}}%
\newcommandx\T[1][usedefault, addprefix=\global, 1=\emptytex]{\mt{T#1}}%
\newcommandx\ee[1][usedefault, addprefix=\global, 1=\emptytex]{\mf{e#1}}%
\global\long\def\where{\cf |}%
\newcommandx\pp[1][usedefault, addprefix=\global, 1=\emptytex]{\mv{p#1}}%
\newcommandx\cc[1][usedefault, addprefix=\global, 1=\emptytex]{\mf{c#1}}%
\newcommandx\MM[1][usedefault, addprefix=\global, 1=\emptytex]{\mf{M#1}}%
\global\long\def\EE{\mf{\mathcal{E}}}%
\newcommandx\ii[1][usedefault, addprefix=\global, 1=\emptytex]{\l{\iota#1}}%
\newcommandx\sub[1][usedefault, addprefix=\global, 1=\emptytex]{\mv{\sigma#1}}%
\global\long\def\TFrac#1#2{
\global\long\def\arraystretch{1.25}%
\genfrac{}{}{0.85pt}{0}{#1}{\vphantom{\ob{T'}}#2}}%
\global\long\def\fsep{\null\ \ \null}%
\global\long\def\ir#1#2{{\displaystyle \TFrac{{\displaystyle \fsep#1\fsep}}{{\displaystyle \fsep#2\fsep}}}}%

\acmConference[META'19]{Workshop on Meta-Programming Techniques and Reflection}{October 20, 2019}{Athens, Greece}{}

\acmYear{2019}

\acmISBN{}

\acmDOI{}

\startPage{1}

\setcopyright{ccby}
\AtBeginDocument{\providecommand\secref[1]{\ref{sec:#1}}}
\AtBeginDocument{\providecommand\subsecref[1]{\ref{subsec:#1}}}
\AtBeginDocument{\providecommand\figref[1]{\ref{fig:#1}}}
\DeclareRobustCommand{\greektext}{%
  \fontencoding{LGR}\selectfont\def\encodingdefault{LGR}}
\DeclareRobustCommand{\textgreek}[1]{\leavevmode{\greektext #1}}

\RS@ifundefined{subsecref}
  {\newref{subsec}{name = \RSsectxt}}
  {}
\RS@ifundefined{thmref}
  {\def\RSthmtxt{theorem~}\newref{thm}{name = \RSthmtxt}}
  {}
\RS@ifundefined{lemref}
  {\def\RSlemtxt{lemma~}\newref{lem}{name = \RSlemtxt}}
  {}

	\newcommand\RuleTitle[1]{%
		{\setlength\fboxrule{\RuleLWidth}\setlength\fboxsep{0pt}%
		\boxed{\ \vphantom{\overline{\strut}}\text{\smash{#1}}\ }}}%

	\newcommand\RuleVSpace{\baselineskip}%
	\newcommand\RuleTSep{0.5\baselineskip}%
	\newcommand\RuleHMin{2em}%
	\newcommand\RuleLWidth{0.4pt}%
	\newcommand\RuleExtraLineSkip{1pt}%

\makeatletter%

	\newcommand\GrowBox[3][0pt]{%
		\setbox0=\hbox{#3}%
		\ht0=\dimexpr\ht0+#2%
		\dp0=\dimexpr\dp0+#1%
		\box0}%

	\newcommand\NewPar{{\setlength{\parskip}{0pt}\vspace{-\RuleExtraLineSkip}\par}\noindent}%

	\newcommand\HVRule@Output[3]{%
		\rule{0pt}{\RuleVSpace}%
		\forQueue{#3}{\@Pos}{%
			\rlap{\hspace{\@Pos}\VRule{\RuleVSpace}}}%
		\rlap{\HRule{#1}\hspace{\dimexpr#2-#1}\HRule{\textwidth-#2}}%
		\NewPar}%

	\newcommand\Equation@Output[1]{%
		\forQueue{#1}{\@EqP}{\@EqP}%
		\NewPar}%

	\newcommand\Start@Output{\NewPar\noindent}%

	\newcommand\HRule[1]{\smash{\rule[-\RuleLWidth/2]{#1}{\RuleLWidth}}}%

	\newlength{\VRH}\newlength{\VRD}%
	\newcommand\VRule[2][0pt]{%
		\setlength{\VRD}{#1}\setlength{\VRH}{#2}%
		\smash{\clap{\rule[-\VRD-\RuleLWidth/2]{\RuleLWidth}{\VRH+\VRD+\RuleLWidth}}}}%

	\newcommand\RuleTitle@Output[2]{%
		\setbox0=\hbox{\RuleTitle@Box{#1}}%
		\edef\tmp{\the\dimexpr#2-(\ht0+\dp0)}%
		\rlap{\raisebox{\tmp}{\box0}}}%

	\newcommand\RuleTitle@OutputVLine[3]{%
		\VRule[#2]{#3}%
		\RuleTitle@Output{#1}{#3}}%

        \newlength{\Title@Depth}%
	\newcommand\RuleTitle@Box[1]{%
		\setbox0=\hbox{\GrowBox{\RuleTSep}{\hbox{\null\hspace{\RuleTSep}#1}}}%
		\setlength{\Title@Depth}{\dp0}%
		\dp0=0pt%
		\raisebox{\Title@Depth}{\box0}}%

	\newcommand\RuleEquation@Output[3]{%
		\GrowBox{#3}{\null\hspace{\dimexpr#2/2}#1\hspace{\dimexpr#2/2}\null}}%

\makeatother%
\makeatletter

\newtoks\@front
\newtoks\@back

\newcommand\@empty@queue{\par}

\newcommand\ifEmptyQueue[1]{\ifx#1\@empty@queue}

\newcommand\enqueue[2]{%
	\@front=\expandafter{#1}\@back=\expandafter{#2}%
	\global\long\edef#1{\the\@front\the\@back\noexpand\par}}

\long\def\@dequeue\par#1\par#2\@dequeue#3#4{\global\long\def#3{\par#2}\global\long\def#4{#1}}

\newcommand\dequeue[2]{\ifEmptyQueue#1\else\expandafter\@dequeue#1\@dequeue#1#2\fi}

\newcommand\forQueue[3]{%
	\ifEmptyQueue#1\else%
		\dequeue{#1}{#2}%
		#3\forQueue{#1}{#2}{#3}%
	\fi}

\newcommand\newQueue[1]{%
	\expandafter\newcommand\csname queue#1\endcsname{\par}%
	\expandafter\newcommand\csname ifEmpty#1\endcsname{%
		\expandafter\ifEmptyQueue\csname queue#1\endcsname}%
	\expandafter\newcommand\csname for#1\endcsname[2]{%
		\expandafter\forQueue\csname queue#1\endcsname{##1}{##2}}%
	\expandafter\newcommand\csname clear#1\endcsname{%
		\expandafter\renewcommand\csname queue#1\endcsname{\par}}%
	\expandafter\newcommand\csname set#1\endcsname[1]{%
		\expandafter\renewcommand\csname queue#1\endcsname{\par##1\par}}%
	\expandafter\newcommand\csname dequeue#1\endcsname[1]{%
		\expandafter\dequeue\csname queue#1\endcsname{##1}}%
	\expandafter\newcommand\csname enqueue#1\endcsname[1]{%
		\expandafter\enqueue\csname queue#1\endcsname{##1}}}

\makeatother
\makeatletter%

\newQueue{RuleInput}%
\newQueue{Pending}%
\newQueue{VLAbove}%
\newQueue{VLCurrent}%

\newcommand\SetMax[2]{%
	\ifdim#1<\dimexpr#2%
	\setlength{#1}{#2}%
	\fi}%

\newcommand\ApplyFirst[2]{\def\First@Apply{#1}\expandafter\First@Arg#2\relax}%
\newcommand\First@Arg[2]{\First@Apply#1\ignorespaces}%

\newcommand\WrapCommand[3]{%
	\expandafter\renewcommand\expandafter#1\expandafter{%
		\expandafter#2\expandafter{#1}#3}}%

\newcount\Eq@Count%
\newlength{\Top@Space}%
\newlength{\Eq@Height}%
\newlength{\Eq@Width}%
\newlength{\Eq@Depth}%
\newlength{\Eq@Space}%
\newcommand\Title@CurrentRight{}%
\newcommand\Title@CurrentLeft{}%
\newcommand\Title@PreviousRight{}%
\newcommand\Default@Left{\textwidth}%

\newcommand\Reset@Vars{%
	\Eq@Count=0\relax%
	\setlength{\Top@Space}{0pt}%
	\setlength{\Eq@Height}{0pt}%
	\setlength{\Eq@Width}{0pt}%
	\setlength{\Eq@Depth}{0pt}%
	\setlength{\Eq@Space}{0pt}%
	\renewcommand\Title@CurrentRight{0pt}%
	\let\Title@CurrentLeft=\Default@Left%
	}%

\newlength{\PR}\newlength{\CL}%
\newcommand\Output@Line{%
	\ifx\Title@PreviousRight\null\else%
		\setlength{\PR}{\Title@PreviousRight}%
		\setlength{\CL}{\Title@CurrentLeft}%
		\let\Lines=\null%
		\expandafter\ifdim\PR>0pt%
			\let\Lines=\relax\fi%
		\expandafter\ifdim\CL<\textwidth%
			\let\Lines=\relax\fi%
		\ifx\Lines\relax%
			\HVRule@Output{\PR}{\CL}{\queueVLAbove}%
		\fi%
	\fi%
	\Equation@Output{\queuePending}%
	\xdef\Title@PreviousRight{\the\dimexpr\Title@CurrentRight}%
	\forVLCurrent{\@Pos}{\edef\@Pos{\@Pos}\enqueueVLAbove{\@Pos}}%
	\Reset@Vars%
	\setlength{\Top@Space}{\RuleVSpace}}%

\newcommand\Compute@Space[2]{%
	\dimexpr(\textwidth-(\Eq@Width+#2))/(\Eq@Count+#1)}%

\newlength{\NewEq@Space}\newbox\Eq@Box%
\newcommand\Prepare@Eq[2]{%
	\setbox\Eq@Box=\hbox{#1}%
	\setlength\NewEq@Space{\Compute@Space{1}{\wd\Eq@Box}}%
	\ifdim\NewEq@Space<\dimexpr\RuleHMin%
		\Output@Line%
	\fi%
	\expandafter\Wrap@Eq\expandafter{\the\dimexpr#2}{#1}%
	\advance\Eq@Count by 1%
	\SetMax\Eq@Height{\ht\Eq@Box+#2}%
	\SetMax\Eq@Depth{\dp\Eq@Box}%
	\setlength{\Eq@Width}{\Eq@Width+\wd\Eq@Box}%
	\setlength{\Eq@Space}{\Compute@Space{0}{0pt}}}%

\newcommand\Wrap@Eq[2]{\WrapCommand{#2}{\RuleEquation@Output}{{\Eq@Space}{#1}}}%

\newcommand\Save@Pos[1]{\expandafter\@Save@Pos\expandafter{\the\dimexpr\Eq@Width}{#1}}%
\newcommand\@Save@Pos[2]{\expandafter\@@Save@Pos\expandafter{\the\Eq@Count}{#1}{#2}}%
\newcommand\@@Save@Pos[3]{\renewcommand#3{\the\dimexpr#2+(\Eq@Space)*#1}}%

\newbox\Title@Box%
\newcommand\@Rules[1]{%
	\Start@Output%
	\clearVLCurrent%
	\setRuleInput{#1}%
	\Reset@Vars%
	\let\Title@PreviousRight=\null%
	\forRuleInput{\@Eq}{%
		\ApplyFirst{\ifx\RuleTitle}{\@Eq}%
			\dequeueRuleInput\@EqR%
			\setbox\Title@Box=\hbox{\RuleTitle@Box{\@Eq}}%
			\ifx\Title@CurrentLeft\Default@Left%
				\Save@Pos{\Title@CurrentLeft}%
			\fi%
			\Save@Pos{\Title@CurrentRight}%
			\Prepare@Eq{\@EqR}{\ht\Title@Box+\dp\Title@Box}%
			\ifnum\Eq@Count>1%
				\WrapCommand{\@Eq}{\RuleTitle@OutputVLine}{{\Eq@Depth}{\Eq@Height}}%
				\enqueueVLCurrent{\Title@CurrentRight}%
			\else%
				\WrapCommand{\@Eq}{\RuleTitle@Output}{\Eq@Height}%
			\fi%
			\enqueuePending{\@Eq}%
			\enqueuePending{\@EqR}%
		\else\ApplyFirst{\ifx\RuleBreak}{\@Eq}%
			\Output@Line%
		\else%
			\Prepare@Eq{\@Eq}{\Top@Space}%
			\enqueuePending{\@Eq}%
		\fi\fi}%
	\Output@Line}%

\newcommand\IRules[1]{%
	\GrowBox[\RuleVSpace]{\RuleVSpace}{%
		\begin{minipage}{\columnwidth}\@Rules{#1}\end{minipage}}}%

\usepackage{environ}%
\NewEnviron{Rules}{\begin{minipage}{\columnwidth}\expandafter\@Rules\expandafter{\BODY}\end{minipage}}%

\usepackage{xspace}
\usepackage[british]{babel}
\makeatletter
\newcommand\ShowInterColumnFrame{
	\usepackage{etoolbox}
	\geometry{showframe}
	\patchcmd{\@outputdblcol}
	  {{\normalcolor\vrule\@width\columnseprule}}
		{\hspace{-6pt}\vrule\@width0.4pt\hspace{6pt}\hfil%
	    {\normalcolor\vrule\@width\columnseprule}%
    	\hfil\rule{6pt}{0pt}\vrule\@width0.4pt\hspace{-6pt}}{}{}}
\makeatother

\definecolor{blue}{HTML}{0000F0} %
\definecolor{purple}{HTML}{700090}
\definecolor{orange}{HTML}{F07000}
\definecolor{teal}{HTML}{0090B0}
\definecolor{brown}{HTML}{A00000} %
\definecolor{green}{HTML}{008000}
\definecolor{pink}{HTML}{F000F0} %

\usepackage{listings}

\makeatletter
\newtoks\@head \newtoks\@tail

\newcommand\@push[2]{%
	\@head=\expandafter{\expandafter\\\expandafter{#2}}\@tail=\expandafter{#1}%
	\long\edef#1{\the\@head\the\@tail}}

\long\def\@@pop\\#1\\#2\@@pop#3#4{\renewcommand#3{\\#2}\def#4{#1}}
\newcommand\@empty@stack{\\}
\newcommand\@pop[2]{\ifx#1\@empty@stack\else\expandafter\@@pop#1\@@pop#1#2\fi}

\newcommand\newstack[1]{%
	\expandafter  \newcommand\csname stack@#1\endcsname{\\}%
	\expandafter  \newcommand\csname    pop#1\endcsname[1]{%
	\expandafter        \@pop\csname stack@#1\endcsname{##1}}%
	\expandafter  \newcommand\csname   push#1\endcsname[1]{%
	\expandafter       \@push\csname stack@#1\endcsname{##1}}}

\def\lst@Activekey#1\@nil@{\let\lst@ifxactive\lst@if\def\lst@active{#1}}
\lst@Key{active}{}{\@ifstar{\lst@true \lst@Activekey}{\lst@false\lst@Activekey}#1\@nil@}
\lst@AddToHook{SelectCharTable}{\ifx\lst@active\@empty\else\expandafter\lst@Active\lst@active{}\relax\fi}
\def\lst@Active#1#2{\ifx\relax#2\@empty\else%
\lst@CArgX #1\relax\lst@CDef{}{\let\lst@next\@empty\lst@ifxactive\lst@ifmode\let\lst@next\lst@CArgEmpty\fi\fi%
\ifx\lst@next\@empty\ifx\lst@OutputBox\@gobble\else\lst@XPrintToken\let\lst@scanmode\lst@scan@m#2%
\fi\let\lst@next\lst@CArgEmptyGobble\fi\lst@next}\@empty\expandafter\lst@Active\fi}

\long\def\get@first#1#2\get@first{\noexpand#1}
\newcommand\getfirst[1]{\expandafter\get@first#1\relax\get@first}

\newcommand\showarg[1]{\def\show@rg{#1}\show\show@rg}
\newcommand\showe[1]{\edef\show@arg{#1}\show\show@arg}

\def\style@other#1#2\style@other{\ifx#1\lst@outputspace\else\apply@style{def}\fi}

\lst@AddToHook{OutputOther}{%
	\edef\@arg{\the\lst@token}%
	\edef\@arg{\getfirst\@arg}%
	\expandafter\style@other\@arg\style@other}
\newcommand\print@token[2]{\lst@token{{#1}}\lst@length#2\relax\let\lst@thestyle\current@style\lst@OutputToken}

\gdef\@style{} \gdef\o@style{} \gdef\e@style{} \newstack{@pars}
\newcommand\clear@style{\global\edef\o@style{\@style}\gdef\@style{}}
\newcommand\force@style[1]{\print@token{}{0}\set@style{#1}\gdef\o@style{#1}}
\newcommand\set@style[1]{\gdef\@style{#1}}
\newcommand\try@style[1]{\ifx\@style\e@style#1\else\use@style\fi}%
\newcommand\use@style{\edef\@s{\@style}\clear@style\expandafter\csname \@s style\endcsname}
\newcommand\current@style{\edef\@s{\@style}\expandafter\csname \@s style\endcsname}
\newcommand\apply@style[1]{\set@style{#1}\use@style}
\newcommand\@open[1]{\push@pars\o@style\apply@style\o@style\print@token{#1}{1}\apply@style{def}}

\newcommand\@close[1]{\pop@pars\o@style\let\oo@style=\o@style\apply@style\o@style\print@token{#1}{1}\apply@style{def}\let\o@style=\oo@style}
\newcommand\@keystyle{\gdef\o@style{key}\keystyle} \newcommand\@litstyle{\gdef\o@style{lit}\litstyle}
\newcommand\@commstyle{\try@style{\apply@style{comm}}}
\newcommand*\id@dispatch{\expandafter\id@switch\the\lst@token\relax}
\def\id@switch#1#2\relax{\try@style{%
	\ifcat|\noexpand#1%
		\apply@style{var}%
		\ifnum`#1<"3A%
			\ifnum`#1>"2F%
				\apply@style{lit}%
			\fi%
		\else%
			\ifnum`#1=\uccode`#1%
				\apply@style{type}%
			\fi%
		\fi%
	\fi}}

\lst@AddToHook{OnEmptyLine}{\vspace{-0.5\baselineskip}}
\makeatother
 
\newcommand{\keystyle}{\color{blue}}

\newcommand{\litstyle}{\color{brown}}

\newcommand{\defActiveMathChar}[2]{%
        \begingroup\lccode`~=`#1\relax%
        \lowercase{\endgroup\def~}{#2}%
        \AtBeginDocument{\mathcode`#1="8000}%
}
\defActiveMathChar{.}{\mspace{-4mu}\texttt{.}\mspace{-2mu}}

\let\omaketitle=\maketitle
\def\maketitle{\let\maketitle=\omaketitle}

\renewcommand\secref[2][]{Section#1 \ref{sec:#2}}
\renewcommand\subsecref[1]{Section \ref{subsec:#1}}

\makeatother

\usepackage{babel}
\usepackage{listings}
\begin{document}
\title{$\name$: The Power of Dynamic Multi-Methods}
\subtitle{(Work-In-Progress Paper)}

\author{Isaac Oscar Gariano}
\affiliation{\institution{Victoria University of Wellington}\city{Wellington}\country{New Zealand}}
\email{Isaac@ecs.vuw.ac.nz}
\author{Marco Servetto}
\affiliation{\institution{Victoria University of Wellington}\city{Wellington}\country{New Zealand}}
\email{Marco.Servetto@ecs.vuw.ac.nz}



\maketitle

\begin{abstract}
Multi-methods are a straightforward extension of traditional (single)
dynamic dispatch, which is the core of most object oriented languages.
With multi-methods, a method call will select an appropriate implementation
based on the values of \emph{multiple} arguments, and not just the
first/receiver. Language support for both single and multiple dispatch
is typically designed to be used in conjunction with other object
oriented features, in particular classes and inheritance. But are
these extra features really necessary?

$\name$ is a dynamic language designed to be as simple as possible
but still supporting flexible abstraction and polymorphism. $\name$
provides only two forms of abstraction: (object) identities and (multi)
methods. In $\name$ method calls are dispatched based on the identity
of arguments, as well as what other methods are defined on them. In
order to keep $\name$s design simple, when multiple method definitions
are applicable, the most \emph{recently defined} one is chosen, not
the most \emph{specific} (as is conventional with dynamic dispatch).

In this paper we show how by defining methods at run-time, we obtain
much of the power of classes and meta object protocols, in particular
the ability to dynamically modify the state and behaviour of `classes'
of objects.
\end{abstract}
\maketitle

\section{Introduction}\label{sec:Introduction}

Most object oriented (and functional\footnote{Since a function or lambda can be thought of as an object with a single
`apply' or `call' method.}) languages support only \emph{single} dynamic dispatch: the code
executed by a method call is determined by a \emph{single} argument
(the receiver). For example, to evaluate a call like $\x\c .\cf{add(\y)}$,
the value of $\x$ will be inspected for a definition of an $\cf{add}$
method (depending on the language, this may be found in a slot/field
\cite{Object-Calculus}, class \cite{Featherweight-Java}, and/or
parent object of $\x$ \cite{Self}) . This approach can be inflexible
for two main reasons: the value of $\y$ is ignored when selecting
the implementation of $\cf{add}$ and only the creator of $\x$ can
define an $\cf{add}$ method.

Multi-methods \cite{Cecil,Clojure,CLOS,Dylan,JavaGI,JPred,Julia,Korz,Multiple-Dispatch}
are one approach to overcoming these limitations: methods can be declared
to dispatch based on the values of \emph{multiple} arguments. Unlike
conventional single dispatch, multi-methods can usually be declared
outside of the objects/classes of any of their arguments. Consider
for example the following Julia code:

\noindent 
\begin{lstlisting}
#add(x `::` Number, y `::` Number) = x + y
#add(x `::` Array, y `::` Array) = 
	[#add(xy[1], xy[2]) for xy = #zip(x, y)]

#add(x `::` Array, y `::` Number) = [#add(xe, y) for xe = x]
#add(x `::` Number, y `::` Array) = [#add(x, ye) for ye = y]

#add([[1, 2], 3], 4) // evaluates to `\c{\sq{\sq{\l 5, \l 6}, \l 7}}`
\end{lstlisting}

Thus a call like $\cf{add\p{\x,\y}}$ will first get the types of
$\x$ \emph{and }$\mv{\y}$, say $\T[{_{1}}]$ and $\T[{_{2}}]$,
and then execute the body of the $\cf{add}$ method defined with $\mf{\p{\c{\_}\cco\T[{_{1}}]\c{,\ }\c{\_}\cco\T[{_{2}}]}}$
(or $\mf{\p{\c{\_}\cco\T[{_{1}^{\prime}}],\ldots,\c{\_}\cco\T[{_{2}^{\prime}}]}}$,
for some super types $\T[{_{1}^{\prime}}]$ and $\T[{_{2}^{\prime}}]$
of $\T[{_{1}}]$ and $\T[{_{2}}]$, respectively).

In the above example, $\ct{Array}$ and $\ct{Number}$ are core language
types, yet we were allowed to define the various $\cf{add}$ methods;
this is in contrast to conventional single dispatch languages, in
which users cannot easily write methods that dynamically dispatch
over pre-defined types/classes.

$\name$ is a language built around multi-methods as an attempt at
being even more flexible by eliminating as many extra concepts as
possible, whilst increasing the flexibility of multi-methods.

\subsection{The $\protect\name$ Programming Language}\label{subsec:Mul}

The language of $\name$\footnote{`$\name$' is pronounced like `mull'; the \textgreek{\textmu}
stands for the ancient Greek word \textgreek{\textmu\'\textepsilon\texttheta\textomicron\textdelta\textomicron\textfinalsigma}
(`methodos'), meaning pursuit of knowledge.} is designed to be as simple as possible, it is an expression based
left-to-right call-by-value language with the following grammar (where
$\x$ and $\m$ are identifiers and $\ii$ is an `identity'):

\vspace{0.5\baselineskip}

$\begin{aligned}\text{(Expression)} & \emptytex & \ee\sop{\Coloneqq} & \gt\mid\newk\mid\defk\ \MM\mid\m[\p{\ob{\ee}}]\\
\emptytex & \emptytex & \emptytex & \ee\sc\ee[']\mid\x\ceq\ee\sc\ee[']\\
\text{(Term)} & \emptytex & \gt\sop{\Coloneqq} & \x\mid\ii\\
\text{(Method)} & \emptytex & \MM\sop{\Coloneqq} & \m[\p{\ob{\pp}\where\ob{\cc}}]\,\br{\ee}\\
\text{(Paramater)} & \emptytex & \pp\sop{\Coloneqq} & \gt\mid\eq\gt\\
\text{(Constraint)} & \emptytex & \cc\sop{\Coloneqq} & \m[\p{\ob{\gt}}]
\end{aligned}
$

\vspace{0.5\baselineskip}

\noindent A \emph{(ground) term} ($\gt$) is either a standard variable
name ($\x$) or an \emph{identity }($\ii$); during reduction, an
$\x$ may be replaced by an $\ii$. \emph{Identities }are the only
kind of value in $\name$, their only intrinsic meaning is whether
they are the same or different (i.e., their `identity'). An $\ii$
cannot appear in the source code, and can only be created by a $\newk$
expression, which will evaluate to a fresh identity that is distinct
from any pre-existing ones.; thus identities cannot be forged.

\emph{Methods} ($\MM$) are global and are dynamically installed by
an expression of the form $\defk\ \m[\p{\ob{\pp}\where\ob{\cc}}]\,\br{\ee}$,
or simply $\defk\ \m[\p{\ob{\pp}}]\,\br{\ee}$ when the $\ob{\cc}$
is empty. Here $\m$ is the name of the method being defined, however
\emph{multiple} methods may be $\defk$ined with the same name; the
$\ob{\pp}$ and $\ob{\cc}$ are (comma separated) lists of parameters
and constraints (respectively); and the body of the method is given
by $\ee$. The $\ee$, $\ob{\pp}$, and $\ob{\cc}$ are interpreted
in the scope where the $\defk$ was evaluated, augmented with appropriate
bindings for any $\pp[{_{i}}]$ of form $\x$: i.e. methods are closures,
as with traditional lambda expressions. A $\defk$ expression does
not reduce to a value/identity, and so cannot be used as an \emph{argument}
to a method call or the RHS of a `$\sceq$', however it can be used
as the \emph{body} of another method\footnote{This restriction is to keep the formalism simple, and we leave it
future work to provide a meaningful value for such an expression.}.

A \emph{method call} expression is of the form $\m[\p{\ob{\ee}}]$
and will first evaluate its arguments to identities $\ob{\ii}$ and
then reduce to $\ee[']\up{\ob{\pp\coloneqq\ii}}$, where $\m[\p{\ob{\pp}\where\ob{\cc}}]\,\br{\ee[']}$
is the \emph{most recently} installed applicable method, and for each
$\pp[{_{i}}]$ of form $\x$, $\up{\ob{\pp\coloneqq\ii}}$ performs
standard capture-avoiding substitution of $\ii$ for $\x$. A method
$\m[\p{\ob{\pp}\where\ob{\cc}}]\,\br{\ee[']}$ is \emph{applicable}
to the call $\m[\p{\ob{\ii}}]$ whenever each $\,\ob{\pp}\up{\ob{\pp\coloneqq\ii}}$
\emph{accepts} the corresponding $\ob{\ii}$, and each $\ob{\cc}\up{\ob{\pp\coloneqq\ii}}$
is \emph{satisfied}. When no such applicable method exists, reduction
will get stuck.

A \emph{parameter} ($\pp$) of form $\gt$ will accept any value,
whereas one of form $\eq\gt$ will only accept \emph{the current}
value of $\gt$. A \emph{constraint }($\cc$) is of form $\m[\p{\ob{\gt}}]$
and is satisfied whenever it has a matching applicable method: i.e.
whenever the call $\m[\p{\ob{\gt}}]$ is defined.

Finally, $\name$ supports standard \emph{sequence} expressions of
the form $\ee\sc\ee[']$, and let expressions of the form $\x\ceq\ee\sc\ee[']$.
We present a formalised set of reduction-rules in Appendix \ref{sec:Appendix}.

The following example demonstrates the aforementioned features in
action:

\noindent 
\begin{lstlisting}
def #foo(n) { n };
// create three distinct identies
bar := new; baz := new; qux := new;
#foo(bar); #foo(baz); #foo(qux); // evaluates to @bar, @baz, and @qux

def #foo(=bar) { baz };
#foo(bar); // Now this evaluates to @baz
#foo(qux); // Still evaluates to @qux, as @qux `$\M{\neq}$` @bar

#foo(qux); // Evaluates to @qux

def #foo(x | #foo2(x)) { #foo2(x) };
#foo(bar); // Still evaluates to @baz, as `\f{foo2\p{\cv{bar}}}` is undefined

def #foo2(=bar) { bar };
#foo(bar); // now returns #foo2(@bar), which returns @bar
\end{lstlisting}

Later on we will introduce syntax sugar for natural numbers, lists,
and lambdas. Our examples also use string literals of the form $\cl "\l{\mathit{text}}\cl "$
which evaluate to an identity $\ensuremath{\ii}$ such that $\cf{print\p{\ii}}$
will print $\l{\ensuremath{\mathit{text}}}$.

We have implemented $\name$\footnote{The source code is available at \url{github.com/IsaacOscar/Mul}.},
with all the aforementioned features, in Racket \cite{Racket} by
using the `brag' \cite{brag} parser generator and an encoding of
our reduction rules in PLT Redex \cite{Redex}. We have used this
to test that all the code examples presented in this paper behave
as indicated.

\section{Abstraction Support}

We now show how $\name$ is powerful enough to encode various language
features including mutable fields, `value equality' (as opposed
to reference equality), multiple dynamic dispatch, and multiple inheritance.
Similarly to meta-object-protocols, $\name$ allows such features
to be used at runtime, allowing state and behaviour to by dynamically
added to whole groups of object a time, thus providing support for
meta-programming. None of this power requires any extensions to $\name$,
the two core abstractions of identities and methods are sufficient;
this is in contrast to most languages were such power is provided
by specific abstractions and syntactic forms.

\subsection{(Mutable) State}\label{subsec:State}

Suppose we want to represent records: collections of named (mutable)
values, we can do this by representing the fields as methods (which
are closures):

\noindent 
\begin{lstlisting}
def #new-point(x, y) { // a 'constructor' for a 'record'
	res := new;
	def #get-x(=res) { x }; // a 'getter'
	def #get-y(=res) { y };
	// @res is like a `\c{\{\v x = \v x, \v y = \v y\}}` record
	res };

point := #new-point(1, 2);
#get-x(point); // returns `\l 1`, like `\c{\v{point}.\v{x} = \l 2}`
\end{lstlisting}

The above works because in $\name$ method definitions, unlike variables
and parameters, have global scope. We can update fields by re-defining
methods (compare this with languages where fields are also methods
\cite{Object-Calculus}):
\begin{lstlisting}
def #get-x(=point) { 2 }; // like `\c{\v{point}.\v x = \l 2}`
#get-x(point); // now returns `\l 2`
\end{lstlisting}

Or, better yet, we can define a setter:

\noindent 
\begin{lstlisting}
// can only call this method if #get-x(@that) is already defined
def #set-x(that, x | #get-x(that)) { 
	def #get-x(=that) { x } };

#set-x(point, 3);
#get-x(point); // now returns `\l 3`
\end{lstlisting}

\noindent This works because in $\name$ it is not an error to redefine
a method: a method call will resolve to the \emph{most recently} defined
applicable method (in this case, the one defined by the most recently
installed $\cf{get-x\p{\eq\cv{point}}}$ method). We could have instead
placed a $\defk\ \cf{set-f1\p{\eq\cv{res},\,\cv x}}$ inside the body
of $\cf{point}$, however the $\cf{set-f1\p{\cv x,\,\cv y\,\where\,\cf{f1\p{\cv x}}}}$
version gives us code reuse: if an identity $\ii$ has a corresponding
$\cf{get-x\p{\ii}}$ method, it will automatically get a $\cf{set-f1\p{\ii,\y}}$
one, even if $\ii$ is not the result of a call to $\cf{new-foo}$.

Note that we can also provide a $\cf{global-set-x\p{\x}}$ method
that will change the value of $\cf{get-x}$ for \emph{all} identities:

\begin{lstlisting}
def #global-set-x(x) {
	def #get-x(that) { x } };

#global-set-x(4);
#get-x(point); // now returns `\l 4`
\end{lstlisting}

This is similar to changing an instance slot to a shared slot using
the meta object protocol of Common Lisp \cite{CLOS}.

\subsection{Value Equality}\label{subsec:Value}

In the previous section we used integer literals, we can obviously
represent these as simply method calls to some `successor' method,
but how do we ensure that a $\cl 1$ appearing somewhere in the source
code is the same identity as another $\cl 1$? Of course we can let
the compiler `intern' these, but there's a more general way, memoise
the successor function:

\noindent 
\begin{lstlisting}
zero := new; // An identity simply representing the number `\c{\l 0}`

def #succ(n) { // returns an identity representing `\c{\v n + \l 1}`
	res := new;
	def #succ(=n) { res }; // memoise the result
	def #pred(=res) { n };
	res };
\end{lstlisting}

\noindent Thanks to the $\defk\ \cf{succ\p{\eq\cv n}}\,\br{\cv{res}}$
line above, any future calls to $\cf{succ}$, with an identical $\cv n$
value will return an identical result. Thus we can safely desugar
a number literal $\l n$ into $\cf{succ}^{\l n}\mf{\p{\cv{zero}}}$.

We can also easily define arithmetic operations such as $\cf{plus}$
in terms of $\cv{zero}$, $\cf{succ}$, and $\cf{pred}$:

\begin{lstlisting}
def #plus(x, =zero) { x };  // Since `\c{\v x + \l 0 = \v x}`
def #plus(x, y | #pred(y)) {
	// Since `\c{\v x + (\v y + \l 1) = (\v x + \v y) + \l 1}`
	// This requires #pred(@y) to be defined (so @y `$\M{\neq}$` `\l 0`)
	#succ(#plus(x, #pred(y))) };

one := #succ(zero);
def #plus(x, =one) { // optimised case for `\c{\v x + \l 1}`
	#succ(x) };
\end{lstlisting}

\noindent Note that the same approach can be used to encode lists
with value equality:

\noindent 
\begin{lstlisting}
empty := new; // The empty list
def #cons(h, t) { // like `\c{\v h : \v t}` in Haskell
	res := new;
	def #cons(=h, =t) { res }; // memoise the result
	def #head(=res) { h }; def #tail(=res) { t };
	res };

list1 := #cons(1, empty); // Or `[\l 1]`
list2 := #cons(1, empty); // Identical to @list1
\end{lstlisting}

Thus we can desugar lists of the form $\sq{\ee[{_{1}}],\ldots,\ee[{_{n}}]}$
into $\c{\f{cons\p{\ee[{_{1}}],\ldots\cf{cons\p{\ee[{_{n}}],\cv{empty}}}\ldots}}}$.

\vspace{-0.3\baselineskip}

\subsection{(Multiple) Dynamic Dispatch}

One of the simplest tools for dynamic dispatch is the lambda expression,
for example consider a classical map function in Haskell:

\noindent 
\begin{lstlisting}
#map(f, [])  = []; `\cM{-- case for empty list}`
#map(f, h `\fixec` t) = f(h) `\fixec` #map(f, t); `\cM{-- case for non empty list}`
#map(\x -> x + 1, [1, 2]); `\c{\M{-- evaluates to} [\l 2, \l 3]}`
\end{lstlisting}

We can encode such dynamic dispatch in $\name$ by using an $\cf{apply}$
method:

\noindent 
\begin{lstlisting}
def #map(f, =empty) { empty };
def #map(f, l | #head(l)) {
	// Here @l `$\M{\neq}$` @empty, since #head(@empty) is undefined
	#cons(#apply(f, #head(l)), #map(f, #tail(l))) };

lam := new; // an identity to represent our lambda expression
def #apply(=lam, x) { #plus(x, 1) }; // the body of the 'lambda'
#map(lam, [1, 2]); // pass the 'lambda'
\end{lstlisting}

We can use this pattern to desugar lambda expressions of the form
$\b{\br{\ob{\pp}\mathrel{\c{=>}}\ee}}$ to `\ensuremath{\pp\ceq\newk\sc}
 $\cf{apply\p{\eq\mv{\v l}\c ,\,\ob{\pp}}\,\br{\ee}}\sc$ $\v l$',
for some fresh identifier $\v l$.

Recall our $\ct{Number}$/$\ct{Array}$ example of multiple dispatch
from \secref{Introduction}, in order to encode this we need to be
able to distinguish between a `natural number' and an `array/list'.
We can do this by using methods:

\noindent 
\begin{lstlisting}
def #as-natural(=zero) { zero };
def #as-natural(n | #pred(n)) { n }; // @n `\c{= \f{succ\p{\ddd}}}`
// #as-natural(@x) is defined iff @x is a 'natural number'

def #as-list(=empty) { empty };
def #as-list(a | #head(a), #tail(a)) { a }; // @a `\c{= \f{cons\p{\ddd}}}`
// #as-list(@x) is defined iff @x is a 'list'
\end{lstlisting}

Note how our $\cf{as-natural}/\cf{as-list}$ methods behave like `structural
type cast' operations: they return their argument if and only if
they have the `structure' of a `number'/`array'; here a `no
applicable method' error would correspond to a failed cast\footnote{This approach has the same problems as structural types: if $\cf{head\p{\mv{\mathit{turtle}}}}$
and $\cf{tail\p{\mv{\mathit{turtle}}}}$ are defined, then so will
$\cf{as-list\p{\mv{\mathit{turtle}}}}$, similarly one can also define
an $\cf{as-list\p{\eq\mv{\mathit{frog}}}}$ method.}. Since we can constrain method definitions by whether such calls
would fail, we can use them to encode multiple dispatch:

\noindent 
\begin{lstlisting}
def #add(x, y | #as-natural(x), #as-natural(y)) {
	#plus(x, y) };
def #add(x, y | #as-list(x), #as-list(y)) {
	// Assuming #zip-map is defined analogously to #map
	#zip-map({xe, ye => #add(xe, ye)}, x, y) };

def #add(x, y | #as-list(x), #as-natural(y)) {
	#map({xe => #add(xe, y)}, x) };
def #add(x, y | #as-natural(x), #as-list(y)) {
	#map({ye => #add(x, ye)}, y) };

#add([[1, 2], 3], 4); // evaluates to `\c{\sq{\sq{\l 5, \l 6}, \l 7}}`
\end{lstlisting}

Thus we still get the power of multiple dispatch without needing a
seperate notion of `type' or `class': methods themselves can play
such a role.

Recall that method definitions apply at \emph{runtime}, and as such,
new methods can be dynamically installed; for example if a programmer
decides at some point that they wish to provide new overloads for
$\cf{add}$ they can do so:

\noindent 
\begin{lstlisting}
def #make-add(x, y, res) {
	def #add(=x, =y) { res } };

#add(1, 1); // returns `\c{\l 2}`
#make-add(1, 1, 3);
#add(1, 1); // now returns `\c{\l 3}`!
\end{lstlisting}

\subsection{(Dynamic) Inheritance}

One common use case for classes is (multiple) inheritance: a derived
class can automatically get all the methods of its base classes. This
is particularly useful as it can significantly reduce code duplication.
We can achieve a similar effect by using a technique like that of
mixins \cite{Mixins}:

\noindent 
\begin{lstlisting}
def #mammal-mixin(self) { // like a 'mammal' class/mixin
	def #as-mammal(=self) { self };
	def #tails(=self) { 1 };
	def #legs(=self) { 4 };
	self };

def #biped-mixin(self) {
	def #as-biped(=self) { self };
	def #legs(=self) { 2 };
	self };

def #kangaroo-mixin(self) {
	#mammal-mixin(self); // inherit #tails and #legs from 'mammal'
	#biped-mixin(self); // override with 'biped's #legs method
	// Note: #biped-mixin(#mammal-mixin(@self)) would also work
	def #as-kangaroo(=self) { self };
	def #arms(=self) { 2 }; // extra method
	self };

k := #kangaroo-mixin(new); // make a new 'kangaroo'
#tails(k); #legs(k); #arms(k); // returns `\l 1`, `\l 2`, `\l 2`
\end{lstlisting}

We can dynamically perform inheritance by passing an existing object
to such a `mixin':

\noindent 
\begin{lstlisting}
mb := #mammal-mixin(new);
#legs(mb); // returns `\l 4`
#biped-mixin(mb); // Almost like adding a new 'class' to @mb
#legs(mb); // now returns `\l 2`
\end{lstlisting}

We can also use our `$\cf{as}$' methods to do something akin to
monkey-patching: we can dynamically install or update a method for
all `instances' of a `mixin':

\noindent 
\begin{lstlisting}
m := #mammal-mixin(new);
#legs(m); #legs(mb); #legs(k); // returns `\l 4`, `\l 2`, and `\l  2`
def #legs(self | #as-biped(self)) { 3 };
#legs(m); #legs(mb); #legs(k); // returns `\l 4`, `\l 3`, and `\l  3`
\end{lstlisting}

\section{Related Work}\label{sec:Related}

Many languages support multi-methods \cite{Multiple-Dispatch}, including
Common Lisp \cite{CLOS}, Clojure \cite{Clojure}, Dylan \cite{Dylan},
Cecil \cite{Cecil}, Julia \cite{Julia}, JavaGI \cite{JavaGI}, JPred
\cite{JPred}, and Korz \cite{Korz}.

Such languages allow methods to specify the required class \cite{CLOS,Dylan,Cecil,JavaGI,Clojure},
type \cite{Julia}, parent object \cite{Korz,Clojure}, and/or a predicate
written in a special purpose sublanguage \cite{JPred}. Unlike $\name$
however, these languages come with complicated dispatch algorithms
in order to determine the `most specific' applicable method to call.
Notably, Common Lisp, Clojure, and Julia also provide support for
runtime redefinition of multi-methods and Common Lisp provides an
$\ck{\p{\k{eql}\ \ee}}$ specifier which is equivalent to $\name$'s
$\eq\x$ (since $\ee$ will be evaluated when the method is defined);
however we have not yet found a language that supports something like
our $\m[\p{\ob{\x}}]$ constraints.

Though most of these languages are large with many features, the language
of Korz has a similar complexity to $\name$: Korz provides `coordinates'
(like $\name$'s identities, except that they can be created with
a parent coordinate) and method definitions. It also supports `dimensions'
(similar to implicit parameters in Scala \cite{Implicits}), the values
of these `dimensions' are implicitly passed throughout all method
calls, their values can be overridden for individual method calls,
and methods can dispatch with respect to them.

\section{Future Work \& Conclusion}\label{sec:Conclusion}

Though the language we presented is both simple and flexible enough
to encode many patterns in object oriented languages, there are however
several notable limitations in $\name$'s design. Firstly, the syntax
can become verbose, especially with the common patterns of $\defk\ \m[\p{\x\where\f{m\p{\x}}}]$
and $\v{res}\ceq\newk\sc$ $\m[\p{\eq\v{res},\ddd}]$ $\ddd\sc$ $\v{res}$;
we are considering various syntax sugars that could alleviate this.

Secondly, $\name$'s expressivity could be improved, such as by adding
higher-order constraints of the form $\f{m\p{\m[\p{\x}]}}$ and providing
support for reflective operations, like dynamically querying defined
methods. Another major limitation is in our dispatch algorithm: a
new method definition will make any pre-existing (less applicable)
methods un-callable. This means that one has to be careful as to what
order they define methods; in addition, one cannot perform something
like a `resend' or `next-method' call, nor can we temporarily
activate methods, as supported by `layer activation' in context-oriented
programming \cite{COP}, thus we are unable to simulate `super'
calls.

Finally, and most importantly, $\name$ is \emph{too} flexible: we
can arbitrarily call define and redefine methods. Not only could this
cause programmers to accidental re-define methods being used by others,
it prevents $\name$ from fully supporting encapsulation. Though $\name$
can prevent arbitrary code from calling `private' methods, it can't
prevent them from being redefined:

\noindent 
\begin{lstlisting}
def #person() {
	res := new; inner := new;
	def #message(=inner) { // a 'private' method
		"Hello World" }; 
	def #speak(=res) { // a 'public' method
		#print(#message(inner)) };
	res };

p := #person();
#speak(p); // Ok, prints `\cl{"Hello World"}`
#message(p); // Error!
def #message(x) { "Goodbye World" };
#speak(p); // Now prints `\cl{"Goodbye World"}`
\end{lstlisting}

Putting method names in protected `packages' or allowing handles
to specific methods (as in Common Lisp) could help, as could `final'
methods. 

Since the easy redefinition of methods makes it very hard to statically
determine what a method call will reduce to, it is likely to be seriously
difficult to implement $\name$ efficiently. We would also like to
extend $\name$ with something akin to a type-system, so that we can
statically ensure the absence of `no applicable method' errors.

\appendix

\section{Reduction Rules}\label{sec:Appendix}

\begin{figure}
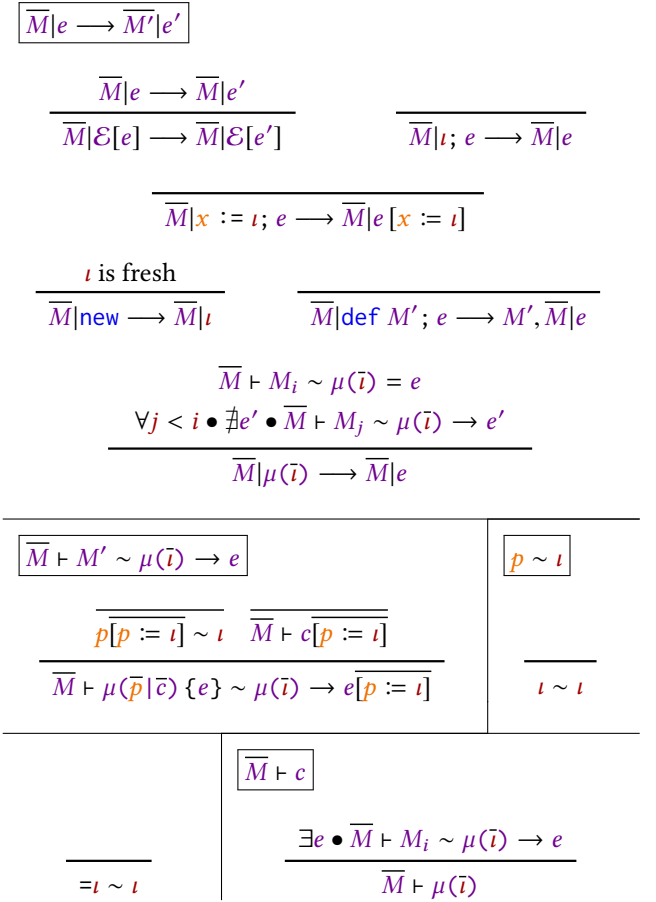

\begin{Rules}
\RuleTitle{$\ob{\MM}|\ee\longrightarrow\ob{\MM[']}|\ee[']$}

$\GrowRule{\ir{\ob{\MM}|\ee\longrightarrow\ob{\MM}|\ee[']}{\ob{\MM}|\EE[\ee]\longrightarrow\ob{\MM}|\EE[\ee[']]}}$

$\ir{\emptytex}{\ob{\MM}|\ii\sc\ee\longrightarrow\ob{\MM}|\ee}$

$\ir{\emptytex}{\ob{\MM}|\x\ceq\ii\sc\ee\longrightarrow\ob{\MM}|\ee\up{\x\coloneqq\ii}}$

$\ir{\ii\text{ is fresh}}{\ob{\MM}|\newk\longrightarrow\ob{\MM}|\ii}$

$\ir{\emptytex}{\ob{\MM}|\defk\ \MM[']\sc\ee\longrightarrow\MM['],\ob{\MM}|\ee}$

$\,\ir{\begin{array}{c}
\ob{\MM}\vdash\MM[{_{i}}]\sim\m[\p{\ob{\ii}}]=\ee\\
\forall\l j<\l i\bullet\nexists\ee[']\bullet\ob{\MM}\vdash\MM[{_{j}}]\sim\m[\p{\ob{\ii}}]\rightarrow\ee[']
\end{array}}{\ob{\MM}|\m[\p{\ob{\ii}}]\longrightarrow\ob{\MM}|\ee}$

\RuleTitle{$\ob{\MM}\vdash\MM[']\sim\m[\p{\ob{\ii}}]\rightarrow\ee$}

$\GrowRule{\ir{\begin{array}{c}
\ob{\pp\ob{\up{\pp\coloneqq\ii}}\sim\ii}\hsep\ob{\ob{\MM}\vdash\cc\ob{\up{\pp\coloneqq\ii}}}\end{array}}{\ob{\MM}\vdash\m[\p{\ob{\pp}\where\ob{\cc}}]\,\br{\ee}\sim\m[\p{\ob{\ii}}]\rightarrow\ee\ob{\up{\pp\coloneqq\ii}}}}$

\RuleTitle{$\pp\sim\ii$}

$\ir{\emptytex}{\ii\sim\ii}$

$\ir{\emptytex}{\eq\ii\sim\ii}$

\RuleTitle{$\ob{\MM}\vdash\cc$}

$\GrowRule{\ir{\exists\ee\bullet\ob{\MM}\vdash\MM[{_{i}}]\sim\m[\p{\ob{\ii}}]\rightarrow\ee}{\ob{\MM}\vdash\m[\p{\ob{\ii}}]}}$
\end{Rules}

\caption{Reduction Rules}\label{fig:Red}
\end{figure}

In \subsecref{Mul} we informally described the semantics of $\name$,
here we provide a formal definition. For reference, here is the grammar
of $\name$ together with a standard left-to-right evaluation context
$\EE$.

\vspace{0.5\baselineskip}

$\begin{aligned}\text{(Expression)} & \emptytex & \ee\sop{\Coloneqq} & \gt\mid\newk\mid\defk\ \MM\mid\m[\p{\ob{\ee}}]\\
\emptytex & \emptytex & \emptytex & \ee\sc\ee[']\mid\x\ceq\ee\sc\ee[']\\
\text{(Term)} & \emptytex & \gt\sop{\Coloneqq} & \x\mid\ii\\
\text{(Method)} & \emptytex & \MM\sop{\Coloneqq} & \m[\p{\ob{\pp}\where\ob{\cc}}]\,\br{\ee}\\
\text{(Paramater)} & \emptytex & \pp\sop{\Coloneqq} & \gt\mid\eq\gt\\
\text{(Constraint)} & \emptytex & \cc\sop{\Coloneqq} & \m[\p{\ob{\gt}}]\\
\text{(Context)} & \emptytex & \EE\sop{\Coloneqq} & \Box\mid\m[\p{\ob{\ii},\EE,\ob{\ee}}]\mid\EE\sc\ee\mid\x\ceq\EE\sc\ee
\end{aligned}
$

\noindent We will use the notation $\ee\up{\pp\coloneqq\ii}$ to perform
a substitution, where $\ee\up{\eq\gt\coloneqq\ii}=\ee$ and $\ee\up{\x\coloneqq\ii}$
performs the usual capture-avoiding substitution; similarly for $\pp[']\up{\pp\coloneqq\ii}$
and $\cc\up{\pp\coloneqq\ii}$.

Our reduction rules are presented in \figref{Red}; our arrow is of
the form $\ob{\MM}|\ee\longrightarrow\ob{\MM[']}|\ee[']$, where the
\emph{state} of the system, $\ob{\MM}$, is the list of installed
methods (ordered by most-recent first) and $\ee$ is the main expression.
We use $\ob{\MM}\vdash\MM[']\sim\m[\p{\ob{\ii}}]\rightarrow\ee$ to
mean that under state $\ob{\MM}$, the method $\MM[']$ is applicable
to the call $\m[\p{\ob{\ii}}]$ and (after substituting in the arguments
$\ob{\ii}$) has body $\ee$; we use $\pp\sim\ii$ to mean that the
parameter $\pp$ accepts the value $\ii$; and $\ob{\MM}\vdash\cc$
to mean that under state $\ob{\MM}$, the constraint $\cc$ is satisfied. 

Note that our reduction rules make no attempt at checking for errors,
in particular reduction will get stuck if a $\defk$ or an unbound
$\x$ is used when an $\ii$ is expected (such as the RHS of a `$\sceq$',
an argument to a method-call, or the RHS of an `$\eq$' parameter).

\bibliographystyle{ACM-Reference-Format}
\bibliography{multi-methods}

\end{document}